\documentclass[aps,pre,twocolumn,galleyproof,superscriptaddress,groupedaddress,showpacs]{revtex4}
\usepackage{graphicx,epsf,dsfont,amssymb,verbatim}

\begin{document}
\title{A coupled order parameter system on a scale-free network}
\author{V. Palchykov}\email[]{palchykov@icmp.lviv.ua}
\affiliation{Institute for Condensed Matter Physics, National
Academy of Sciences of Ukraine, UA--79011 Lviv, Ukraine}
\author{C.\ von Ferber}
\email[]{C.vonFerber@coventry.ac.uk}
 \affiliation{Applied Mathematics Research Centre, Coventry University,
Coventry CV1 5FB, UK}
 \affiliation{Physikalisches Institut,
Universit\"at Freiburg, D-79104 Freiburg, Germany}
\author{R. Folk}\email[]{folk@tphys.uni-linz.ac.at}
 \affiliation{Institut f\"ur Theoretische Physik, Johannes Kepler
Universit\"at Linz, A-4040, Linz, Austria}
\author{Yu. Holovatch}\email[]{hol@icmp.lviv.ua}
\affiliation{Institute for Condensed Matter Physics, National
Academy of Sciences of Ukraine, UA--79011 Lviv, Ukraine}
\affiliation{Institut f\"ur Theoretische Physik, Johannes Kepler
Universit\"at Linz, A-4040, Linz, Austria}
\date{March 27, 2009}
\begin{abstract}
The system of two scalar order parameters on a complex scale-free
network is analyzed in the spirit of Landau theory. To add a
microscopic background to the phenomenological approach we also
study a particular spin Hamiltonian that leads to coupled scalar
order behavior using the mean field approximation. Our results show
that the system is characterized by either of two types of ordering:
either one of the two order parameters is zero or both are non-zero
but have the same value. While the critical exponents do not differ
from those of a model with a single order parameter on a scale free
network there are notable differences for the amplitude ratios and
susceptibilities. Another peculiarity of the model is that the
transverse susceptibility is divergent at all $T<T_c$, when $O(n)$
symmetry is present. This behavior  is related to the appearance of
Goldstone modes.
\end{abstract}
 \maketitle
\section{Introduction}
The topology of many natural and man-made networks (social networks,
biological, technological and transportation systems) strongly
differs from the topology of regular lattices or even random graphs.
Often these networks show scale-free behavior
\cite{Albert02,Dorogovtsev02a,Newman03a} -- the probability of a
randomly chosen node to have a degree $k$ (to have $k$ links)
follows a power law
\begin{equation}\label{eq01}
P(k) = A k^{-\lambda}.
\end{equation}
Other integral parts of many real networks are a small-world effect
and high clustering, resulting in specific features of cooperative
phenomena on such systems. This has sparked interest in the analysis
of different spin models on complex networks \cite{Dorogovtsev08}.
Moreover, such models have interesting applications. For example,
the opinion of each individual of a social network may be
represented by an Ising spin e.g. for simple YES or NO alternatives.
Such a model may describe phenomenon of opinion formation, namely,
individuals change their mind under the influence of their
acquaintances \cite{soc}.

The properties of the order-disorder phase transition of the Ising
model on complex networks strongly depend on the node degree
distribution (\ref{eq01}). Numerical simulations \cite{Aleksiejuk02}
and analytical calculations \cite{Bianconi02} of the Ising model on
Barabasi-Albert scale-free networks ($\lambda=3$) as well as
different analytical approaches \cite{Leone02,Dorogovtsev02b} and
Monte Carlo simulations \cite{Herrero04} for the Ising model on
networks with arbitrary degree distributions have been performed.
Three types of behavior were found depending on the respective
behavior of the moments $\langle{k^2}\rangle$ and
$\langle{k^4}\rangle$ of the degree distribution, which are related
to the value of the $\lambda$ exponent. Namely, if
$\langle{k^4}\rangle$ and $\langle{k^2}\rangle$ are finite
($\lambda>5$), the behavior of the system is described by the
standard mean-field critical exponents. If $\langle{k^4}\rangle$
diverges and $\langle{k^2}\rangle$ is finite, which corresponds to
$3<\lambda\leq5$, the critical behavior is governed by either mean
fields exponents with logarithmic corrections ($\lambda=5$) or by
nontrivial $\lambda$-dependent critical exponents. Finally, if both
$\langle{k^4}\rangle$ and $\langle{k^2}\rangle$ diverge
($2<\lambda\leq3$), the critical temperature becomes divergent (for
the infinite-size networks) and the system is always ordered.
Furthermore, other models on scale free networks, namely the XY
\cite{Kwak07} and Potts \cite{Igloi02,Dorogovtsev04} models (for a
more detailed list see e.g. \cite{Holovatch06,Dorogovtsev08}) also
show peculiarities depending on the value of $\lambda$.

Rather recently, critical phenomena on complex networks have been
studied in the spirit of Landau theory \cite{Goltsev03}. The power
of the latter is that it is independent of the origin of the
interactions between the particles, and therefore it may be applied
to a wide range of systems. The main new feature of the
phenomenological theory of critical phenomena on complex networks,
that differs from the standard Landau theory is the dependence of
the coefficients on the moments $\langle{k^i}\rangle$ of the degree
distribution (\ref{eq01}).

Landau theory for two interacting scalar order parameters is widely
used to analyze systems with several possible types of ordering
(e.g. ferromagnetic and antiferromagnetic, ferroelectric and
ferromagnetic, or structural and magnetic ordering). Such
combination of order parameters may be described by a model of two
scalar order parameters $x_1,x_2$, which are coupled
\cite{Imry75,Watanabe85}. Assuming that the Landau free energy is
analytic and symmetric with respect to the signs of $x_1$ and $x_2$
the lowest order coupling is biquadratic
\begin{equation}\label{eq02}
\Phi(\vec{x},T) = \frac{a}{2}(T-T_c)|\vec{x}|^2 +
\frac{b}{4}|\vec{x}|^4 + \frac{c}{4}x_1^2x_2^2.
\end{equation}
where $\vec{x}=(x_1,x_2)$, $|\vec{x}|^2=x_1^2+x_2^2$, $a$, $b$, $c$
are phenomenological Landau parameters, $T$ and $T_c$ are
temperature and critical temperature, correspondingly. A possible
application of this model of  coupled order parameters on a social
network may reflect the coupling between the preferences for a
candidate and a party in an election (or similar scenarios)
\cite{note1}. The free energy (\ref{eq02}) corresponds to the free
energy of an $n$-vector anisotropic cubic model in the case $n=2$.
The latter is obtained from the $O(n)$ invariant free energy by
adding invariants of the symmetry group $B_n$ of the $n$-dimensional
hypercube \cite{cubic model}.

The aim of our work is to generalize the Landau theory for models on
complex networks \cite{Goltsev03} to the case of two interacting
order parameters with a free energy symmetry given by (\ref{eq02}).
The structure of our paper is as follows. The next section
(\ref{II}) lays out the basic assumptions of the theory and the
peculiarities of the free energy construction  and compares the
approach with a microscopic model. Section \ref{III} describes the
stable states and the phase diagrams of the system. The behavior of
the thermodynamic functions, isothermal susceptibilities and the
heat capacity is described in section \ref{IV}. We conclude with an
outlook in section \ref{V}. Some details of our calculations are
given in appendices \ref{A} and \ref{B}.

\section{Free energy}\label{II}

This section is devoted to the construction of a generalized Landau
theory for a system with two coupled order parameters on a network
(\ref{IIa}). Besides, we derive a corresponding free energy starting
with a microscopic spin Hamiltonian and compare both approaches
(\ref{IIb}).

\subsection{Generalized Landau theory}\label{IIa}
In the spirit of the Landau approach we assume that the system may
display some ordering which can be quantitatively characterized by
two order parameters $x_1$ and $x_2$. For convenience let us
introduce a vector $\vec{x}=(x_1,x_2)$. Following the work of Ref.
\cite{Goltsev03} we assume that the Landau free energy is not only a
function of the order parameters, conjugated field $\vec{h}$, and
temperature  but also depends on the node degree distribution $P(k)$
\begin{equation}\label{eq04}
\Phi (\vec{x},T,\vec{h}) = \int_1^{k_{max}}{dk P(k) f(\vec{x},
k\vec{x})} \, - \, \vec{h}\vec{x},
\end{equation}
where $f(\vec{x}, k\vec{x})$ represents the contribution to the free
energy of an individual node of degree $k$, and $k_{max}$ is the
maximal node degree of the network. Note, that $k_{max}\to\infty$ is
implied for an infinite size system with a power-law node degree
distribution as in Eq. (\ref{eq01}). That $f(\vec{x}, k\vec{x})$
depends not only on the order parameters $x_1$ and $x_2$, but also
on $k\vec{x}$ may be understood by simple reasoning. It reflects
that any node with $k$ neighbors is subject to a field $k\vec{x}$ of
these neighbors.

The next basic assumption in the case of a scalar order parameter
$x$ is that $f(x, kx)$ is an analytic function of $x$ and $kx$
\cite{Goltsev03}. In the case of two order parameters we assume that
$f(\vec{x},k\vec{x})$ is now an analytic function of $x_1$, $x_2$,
$kx_1$ and $kx_2$ and may be represented as a series in their powers
\begin{equation}\label{eq05}
f(\vec{x},k\vec{x})=\sum_{l_1,l_2,m_1,m_2=0}^{\infty}f_{l_1l_2m_1m_2}x_1^{l_1}x_2^{l_2}(kx_1)^{m_1}(kx_2)^{m_2},
\end{equation}
where $f_{l_1l_2m_1m_2}$ are functions which in general may depend
on the temperature $T$ and an external field $\vec{h}$. Moreover,
some relations between these coefficients are implied by the
symmetry of the system as described by equation (\ref{eq02}). In
this case the function $f(\vec{x}, k\vec{x})$ may be represented as
\begin{eqnarray}\label{eq06}
f(\vec{x}, k\vec{x}) &=& f_0 + \sum_{i=0}^{2}a_ik^i|\vec{x}|^2 +
\sum_{i=0}^{4}b_ik^i|\vec{x}|^4 \\ \nonumber &+&
\sum_{i=0}^{4}c_ik^i\sum_{\mu=1}^2x_{\mu}^4+\ldots
\end{eqnarray}
where $f_0$, $a_i$, $b_i$, $c_i$ are convenient notations for the
coefficients $f_{l_1l_2m_1m_2}$.

Naturally, the free energy (\ref{eq04}) must be finite if the order
parameters are finite. This condition is satisfied in particular if
the behavior of the function $f(\vec{x}, k\vec{x})$ at large
$k|\vec{x}| \to \infty$ is bounded by
\begin{equation}\label{eq07}
f(\vec{x}, k\vec{x}) \sim k|\vec{x}|,\hspace{1em}k|\vec{x}| \to
\infty.
\end{equation}

The assumptions (\ref{eq04}) and (\ref{eq06}) as well as  condition
(\ref{eq07}) serve as our basis to analyze the  phase transitions in
the coupled order parameter system following the standard approach
of Landau theory \cite{Landau}.

Substituting (\ref{eq06}) into (\ref{eq04}) and taking into account
that the coefficient of $|\vec{x}|^2$ in the free energy is equal to
zero at the critical point, the equation for the critical
temperature $T_c$ as function of the moments of the degree
distribution is found in the same manner as in the case of a scalar
order parameter \cite{Goltsev03} to be
\begin{equation}\label{eq08}
a_0(T_c)+a_1(T_c)\langle{k}\rangle+a_2(T_c)\langle{k^2}\rangle=0.
\end{equation}
If $a_0(T_c)=0$, the critical temperature is a function of
$\langle{k^2}\rangle/\langle{k}\rangle$. This statement is in
accordance with the exact result for the Ising model on networks
obtained analytically \cite{Leone02,Dorogovtsev02b} and confirmed
numerically \cite{Herrero04}, where $T_c$ follows
\begin{equation}\label{eq09}
\frac{1}{T_c}=\frac{1}{2}\ln\Big(\frac{\langle{k^2}\rangle}{\langle{k^2}\rangle-2\langle{k}\rangle}\Big).
\end{equation}

Before we embark to calculate the free energy let us discuss an
essential point that is the origin of many of the peculiarities of
cooperative phenomena on networks. For scale-free networks with a
node degree distribution as in Eq. (\ref{eq01}) one finds in general
that all moments $\langle{k^i}\rangle$ with $i<\lambda-1$ are
finite, whereas all moments with $i\geq\lambda-1$ diverge. If we
restrict the series in Eq. (\ref{eq06}) to the fourth power of the
order parameter, there are no relevant divergent moments for
$\lambda>5$. Nevertheless, if $\langle{k^4}\rangle$ or lower moments
of the degree distribution are divergent ($\lambda\leq5$), as often
found for real networks, the free energy (\ref{eq04}) at the first
sight may seem to be infinite for any nonzero values of the order
parameters, a behavior which certainly is unphysical. In fact, the
correct way to calculate the free energy is to take into account all
the orders of the series (\ref{eq06}). This procedure ensures to a
behavior of the function $f(\vec{x}, k\vec{x})$ at large values of
$k|\vec{x}| \to\infty$ as described by equation (\ref{eq07}).
Therefore, we collect all terms in equation (\ref{eq06}) containing
$k^i$ with $i\geq\lambda-1$ together with the highest orders of the
series (\ref{eq06}) in a function $g(\vec{x}, k\vec{x})$:
\begin{eqnarray}\label{eq10}
f(\vec{x}, k\vec{x}) &=& f_0 + \sum_{i=0}^{i_2}a_ik^i|\vec{x}|^2 +
\sum_{i=0}^{i_4}b_ik^i|\vec{x}|^4\\\nonumber
&+&\sum_{i=0}^{i_4}c_ik^i\sum_{\mu=1}^2x_{\mu}^4+g(\vec{x},
k\vec{x}).
\end{eqnarray}
Here $i_2$ is the maximal integer that satisfies both conditions
$i_2\leq2$ and $i_2<\lambda-1$. Respectively, $i_4$ is the maximal
integer that satisfies both $i_4\leq4$ and $i_4<\lambda-1$. Now it
is straight forward easy to integrate the part of $f(\vec{x},
k\vec{x})$ that does not include $g(\vec{x}, k\vec{x})$. Any
peculiarities are connected with the integration of $g(\vec{x},
k\vec{x})$. Let us therefore investigate the properties of this
function. Comparing (\ref{eq10}) with (\ref{eq06}) one finds that
for small values of $k|\vec{x}|$ and for $3<\lambda\leq5$ this
function behaves as
\begin{equation}\label{eq11}
g(\vec{x}, k\vec{x}) = b_4(k|\vec{x}|)^4 + c_4 \frac{x_1^4 +
x_2^4}{|\vec{x}|^4}(k|\vec{x}|)^4, \hspace{1ex} k|\vec{x}|\to0.
\end{equation}
For $2<\lambda\leq3$ one finds the following behavior:
\begin{eqnarray}\nonumber
g(\vec{x}, k\vec{x}) &=& a_2(k|\vec{x}|)^2 + \big(b_2 +
c_2\frac{x_1^4 + x_2^4}{|\vec{x}|^4}
\big)|\vec{x}|^2(k|\vec{x}|)^2\\\label{eq12} &+& \big(b_3 +
c_3\frac{x_1^4 +
x_2^4}{|\vec{x}|^4}\big)|\vec{x}|(k|\vec{x}|)^3\\\nonumber &+&
\big(b_4 + c_4\frac{x_1^4 + x_2^4}{|\vec{x}|^4}\big)(k|\vec{x}|)^4,
\hspace{1ex} k|\vec{x}|\to0.
\end{eqnarray}
We do not consider the case $\lambda\leq2$ here as far as
$\langle{k}\rangle$ is not defined.

In order to satisfy condition (\ref{eq07}) for a finite free energy,
the behavior of $g(\vec{x}, k\vec{x})$ is restricted for large
values of $k|\vec{x}| \to\infty$ by the highest explicitly written
term of $f(\vec{x}, k\vec{x})$ in (\ref{eq10}). Namely, for
$k|\vec{x}|\to\infty$  $g(\vec{x},k\vec{x})$ is restricted by
 \begin{equation}\label{eq12a}
g(\vec{x},k\vec{x})\sim\left\{\begin{array}{ll}
 (k|\vec{x}|)^{3}, & 4<\lambda\leq 5\\
 (k|\vec{x}|)^{2}, & 3<\lambda\leq 4\\
 (k|\vec{x}|),     & 2<\lambda\leq 3.\\
 \end{array}\right.
\end{equation}
To perform the integration of $g(\vec{x}, k\vec{x})$ in
(\ref{eq04}), note that it actually depends on $\vec{x}$ and
$k|\vec{x}|$, $g(\vec{x}, k\vec{x})\equiv g(\vec{x}, k|\vec{x}|)$
(see (\ref{eq11}) - (\ref{eq12a})). Let us pass to a new variable
$y=k|\vec{x}|$, which ranges from $|\vec{x}|$ to infinity for
infinite size networks. For a network with a power law node degree
distribution (\ref{eq01}) one may then write
\begin{equation}\label{app01}
\int_1^{\infty}dk P(k) g(\vec{x}, k|\vec{x}|) =
A|\vec{x}|^{\lambda-1}\int_{|\vec{x}|}^{\infty}\frac{dy}{y^{\lambda}}
g(\vec{x}, y).
\end{equation}
As only the asymptotics of $g(\vec{x}, y)$ are fixed, let us write
\begin{equation}\label{app04}
\int_{|\vec{x}|}^{\infty}\frac{dy}{y^{\lambda}}g(\vec{x},y) =
\int_{\varepsilon}^{\infty}\frac{dy}{y^{\lambda}}g(\vec{x},y) -
\int_{\varepsilon}^{|\vec{x}|}\frac{dy}{y^{\lambda}}g(\vec{x},y),
\end{equation}
where $\varepsilon$ is a small positive number
$0<\varepsilon<|\vec{x}|$; obviously, both sides of this expression
do not depend on $\varepsilon$. The first term on its right-hand
side is convergent, due to the asymptotic behavior (\ref{eq12a}). In
the second term $g(\vec{x}, y)$ may be replaced by its expansion for
small values of $y$ (\ref{eq11}), (\ref{eq12}).

Following this procedure, one may obtain the free energy. Details of
the integration of the expression (\ref{app04}) for the case
$4<\lambda<5$ are given in Appendix \ref{A} (analogue calculations
can be performed for other values of $\lambda$). In the following,
we will consider zero external magnetic field $\vec{h}=0$ and drop
the explicit $\vec{h}$-dependence from our notations. Let us present
the resulting expressions for the Landau free energy for different
ranges of values of $\lambda$. We treat the cases ({\em a})
$\lambda>5$, ({\em b}) $\lambda=5$, ({\em c}) $3<\lambda<5$, ({\em
d}) $\lambda=3$ and ({\em e}) $2<\lambda<3$. As we will see,
differences between usual Landau theory and that on a scale-free
network become apparent starting from the marginal case $\lambda=5$.

({\em a}) {\em Case $\lambda>5$}

In this case the free energy may be found easily by substituting
(\ref{eq06}) into (\ref{eq04}) and performing the integration. The
free energy reads:
\begin{equation}\label{feg5}
\Phi(\vec{x},T) = f_0 + \frac{a}{2}(T-T_c)|\vec{x}|^2 +
\frac{b^{(\lambda)}}{4}|\vec{x}|^4 +
\frac{c^{(\lambda)}}{4}x_1^2x_2^2.
\end{equation}
The specific network properties are expressed by the coefficients:
\begin{equation}\label{eq002}
\frac{a}{2}(T-T_c) = a_1\langle{k}\rangle + a_2\langle{k^2}\rangle
\end{equation}
\begin{equation}\label{eq003}
b^{(\lambda)}=4b_4\langle{k^4}\rangle,\hspace{1em}
c^{(\lambda)}=4c_4\langle{k^4}\rangle.
\end{equation}
As seen below, (\ref{eq002}) also holds for $3<\lambda\leq5$.

({\em b}) {\em Case $\lambda=5$}

In this case the free energy reads:
\begin{eqnarray} \nonumber
\Phi(\vec{x},T) &=& f_0 + \frac{a}{2}(T-T_c)|\vec{x}|^2 +
\frac{b^{(\lambda)}}{4}|\vec{x}|^4\ln\frac{1}{|\vec{x}|} \\
 & + & \label{fe5}
\frac{c^{(\lambda)}}{4}x_1^2x_2^2\ln\frac{1}{|\vec{x}|}.
\end{eqnarray}
In this marginal case the free energy displays logarithmic
corrections to the standard mean-field behavior. The coefficient
$a(T-T_c)$ is described by (\ref{eq002}) and the other coefficients
are as follows
\begin{equation}\label{eq005}
b^{(\lambda)}=4A(b_4+c_4),\hspace{1em} c^{(\lambda)}=-8Ac_4.
\end{equation}

({\em c}) {\em Case $3<\lambda<5$}

Here, the free energy reads:
\begin{equation}\label{fe35}
\Phi(\vec{x},T) = f_0 + \frac{a}{2}(T-T_c)|\vec{x}|^2 +
\frac{b^{(\lambda)}}{4}|\vec{x}|^{\lambda-1} +
\frac{c^{(\lambda)}}{4}\frac{x_1^2x_2^2}{|\vec{x}|^4}|\vec{x}|^{\lambda-1}.
\end{equation}
In this case the free energy (\ref{fe35}) explicitly depends on
$\lambda$. The coefficient $a(T-T_c)$ is also described by
(\ref{eq002}), whereas to get expressions for $b^{(\lambda)}$ and
$c^{(\lambda)}$ from the integration of $g(\vec{x}, k\vec{x})$, one
needs to perform explicit calculations in parallel to those,
presented in the Appendix \ref{A}.

({\em d}) {\em Case $\lambda=3$}

Here, the free energy reads:
\begin{equation}\label{fe3}
\Phi(\vec{x},T) = f_0 + C|\vec{x}|^2 -
D|\vec{x}|^2\ln\frac{1}{|\vec{x}|} +
E\frac{x_1^2x_2^2}{|\vec{x}|^4}|\vec{x}|^2.
\end{equation}

({\em e}) {\em Case $2<\lambda<3$}

In this case we find a free energy of the form
\begin{equation}\label{fe23}
\Phi(\vec{x},T) = f_0 + C^{'}|\vec{x}|^2 +
D^{'}|\vec{x}|^{\lambda-1} +
E^{'}\frac{x_1^2x_2^2}{|\vec{x}|^4}|\vec{x}|^{\lambda-1}.
\end{equation}

For the cases ({\em d}) and ({\em e}) we give explicitly only the
expressions for $D$ (for $\lambda=3$) and $C^{'}$ (for
$2<\lambda<3$):
\begin{equation}\label{eq006}
D=-Aa_2,\hspace{1em}C^{'} = a_1\langle{k}\rangle -
\frac{Aa_2}{3-\lambda}.
\end{equation}
For an example how to calculate the other coefficients from the
integration of $g(\vec{x}, k\vec{x})$, see Appendix \ref{A}.

Note, that for $2<\lambda\leq3$, the term of order $|\vec{x}|^2$ is
no more the leading one. Terms of lower order of magnitude become
relevant. In particular there is a term
$|\vec{x}|^2\ln|\vec{x}|^{-1}$ for $\lambda=3$, and a term
$|\vec{x}|^{\lambda-1}$ for $2<\lambda<3$.

Before passing to the details of the phase diagram that results from
the expressions for the Landau free energy (\ref{feg5}),
(\ref{fe5}), (\ref{fe35}), (\ref{fe3}), (\ref{fe23}), we first
proceed to show that the Landau free energy may also be derived from
a spin system on a network by calculating its partition function in
the simplest of approximations.

\subsection{Anisotropic Hamiltonian}\label{IIb}

One of the ways to get the Landau free energy with two coupled
scalar order parameters is to start from two coupled spin subsystems
\cite{note2}. Another way is to consider a single spin system with a
cubic anisotropy term. Let us here use the second option,
considering a spin model on a complex network described by a
Hamiltonian with an anisotropic term
\begin{equation}\label{eq13}
H = -J\sum_{\langle i,j \rangle}\vec{s_i}\cdot\vec{s_j} +
u\sum_{i=1}^{N}\sum_{\nu=1}^{2}s_{\nu,i}^4
\end{equation}
where $\vec{s_i}$ and $\vec{s_j}$ are spins on nodes $i$ and $j$
correspondingly, $J$ and $u$ are the coupling and anisotropy
constants, the notation $\sum_{\langle i,j \rangle}$ denotes the
summation over all pairs of connected nodes, the index $\nu$ numbers
the components of the two-component vector, $\vec{s_i}\cdot\vec{s_j}
= \sum_{\nu=1}^2 s_{\nu,i} s_{\nu,j}$ is a scalar product. Again, as
above we will consider the case when the network node degree
distribution obeys a power law decay (\ref{eq01}). Note, that the
Hamiltonian (\ref{eq13}) represents an $n$-vector anisotropic cubic
model \cite{cubic model} in the case $n=2$.

Here, we consider the Hamiltonian (\ref{eq13}) in the spirit of a
mean-field approach. Applying the mean field approach to a model
that is defined on a regular lattice (equal degree $k$ for all
nodes), each node is characterized by the same mean spin
$\langle{\vec{s}}\rangle$ and experiences the effective field
$\langle{k}\rangle \langle{\vec{s}}\rangle$ of its
$\langle{k}\rangle$ neighbors. In the case of a complex network,
this assumption may be applied only to nodes with the same degree:
in the simplest approximation each $k$-degree node experiences the
same mean spin $\langle{\vec{s}}\rangle_k$. In turn, the mean spin
value per node $\langle\vec{s}\rangle$ may be expressed in terms of
$\langle{\vec{s}}\rangle_k$ as
\begin{equation}\label{eq13a}
\langle\vec{s}\rangle = \sum_k P(k)\langle\vec{s}\rangle_k.
\end{equation}
On the other hand, it can be found from the thermodynamical
definition
\begin{equation}\label{eq13b}
\langle\vec{s}\rangle = - \Big (\frac{\partial G(T,
\vec{h})}{\partial \vec{h}}\Big )_T.
\end{equation}
Here $G(T, \vec{h})$ is the appropriate thermodynamical potential
and $\vec{h}$ is an external field.

Node $i$ experiences the effective field of its $k_i$ neighbors.
This field may be quantitatively characterized by the mean value
$\vec{\sigma^{i}}$ of the spins surrounding the $i$-th node
\cite{Leone02,Dorogovtsev02b}:
\begin{equation}
\vec{\sigma^{i}} = \frac{1}{k_i}\sum_{\langle j\rangle}\vec{s_j}.
\end{equation}
Here the sum over $j$ spans the $k_i$ nearest neighbors of node $i$.
Now, in the spirit of the mean field theory one assumes that
$\vec{\sigma^{i}}$ does not depend on the node number $i$
\begin{equation}
\vec{\sigma^{i}} = \vec{\sigma},\hspace{2em}i=1{\ldots}N.
\end{equation}
Note, that the above defined value $\vec{\sigma}$ differs from the
mean spin value per node $\langle\vec{s}\rangle$. Equation
(\ref{eq13b}) gives us the relation between the mean spin
$\langle\vec{s}\rangle$ and the effective spin $\vec{\sigma}$ per
neighbor and may be treated as a self-consistency equation.

To proceed with the Hamiltonian (\ref{eq13}), we introduce the
deviation of every spin component $s_{\nu,i}$ from the corresponding
component of average spin per neighbor $\vec{\sigma}$:
\begin{equation}\label{eq14}
\Delta s_{\nu,i} = s_{\nu,i} - \sigma_\nu.
\end{equation}
Substituting $s_{\nu,i} = \sigma_\nu + \Delta s_{\nu,i}$ into the
scalar product in (\ref{eq13}) and neglecting the terms of order
$O((\Delta s)^2)$ we arrive at the mean-field Hamiltonian:
\begin{equation}\label{eq15}
H_{MF} = \sum_{i=1}^N H_{MF}^i
\end{equation}
with
\begin{equation}\label{eq15a}
H_{MF}^i = \frac{1}{2}J\langle{k}\rangle \sigma^2 -
Jk_i\sum_{\nu=1}^2 \sigma_{\nu} s_{\nu,i} +
u\sum_{\nu=1}^2s_{\nu,i}^4.
\end{equation}
Here, $\sigma^2 = \sigma_1^2 + \sigma_2^2$. Now, the partition
function is reduced to a product of single-site traces:
\begin{equation}\label{eq15b}
Z_{MF} = \prod_{i=1}^N {\rm Tr}_i\, e^{-H_{MF}^i/T}.
\end{equation}
Here, the single-site trace ${\rm Tr}_i(\ldots)$ denotes the
integration over all possible directions of $\vec{s_i}$:
\begin{equation}\label{eq15c}
{\rm Tr}_i(\ldots) = \int d\vec{s_i}\delta(L-|\vec{s_i}|)(\ldots).
\end{equation}
The $\delta$-function ensures that all spins $\vec{s_i}$ have the
same absolute value $L$. Substituting (\ref{eq15a}) into
(\ref{eq15b}) and taking the trace (some details of the calculations
are given in Appendix \ref{B}) one arrives at the free energy per
site:
\begin{equation}\label{eq15da}
F (\vec{\sigma},T)  = -T/N \ln Z_{MF}.
\end{equation}
As usual in the mean field approach, the free energy (\ref{eq15da})
depends on the macroscopic mean field variable $\vec{\sigma}$. The
last is to be eliminated by corresponding minimization of $F
(\vec{\sigma},T)$. The  expression for the free energy per site
reads:
\begin{equation}\label{eq15d}
F (\vec{\sigma},T)  = \frac{1}{N}\sum_{i=1}^N \hat{f}(\vec{\sigma},
k_i\vec{\sigma})
\end{equation}
with
\begin{eqnarray}\label{eq18}
\hat{f}(\vec{\sigma}, k_i\vec{\sigma}) &=& - T\ln\big( 2\pi L\big)
+\frac{1}{2}\frac{T^2}{JL^2}k_i|\vec{\xi}|^2
\\\nonumber
&&-T\ln\Big( I_{0}(k_i|\vec{\xi}|) - \frac{uL^4}{T}\Big[
6\frac{I_{2}(k_i|\vec{\xi}|)}{(k_i|\vec{\xi}|)^2}\\\nonumber
&&+6\frac{I_{3}(k_i|\vec{\xi}|)}{k_i|\vec{\xi}|} +
\frac{\sum_{\nu=1}^2 \xi_{\nu}^4}{|\vec{\xi}|^4}
I_{4}(k_i|\vec{\xi}|) \Big] \Big)
\end{eqnarray}
and
\begin{equation}\label{eq17}
\vec{\xi} = \frac{JL}{T}\vec{\sigma}.
\end{equation}

In (\ref{eq18}), $I_n(z)$ are modified Bessel functions
\cite{Abramowitz} of the first kind:
\begin{equation}\label{bes}
I_n(z) = \frac{1}{2\pi {\rm i}}\oint
e^{(z/2)(\omega+1/\omega)}\omega^{-n-1} {\rm d}\omega.
\end{equation}
It is instructive to observe that in (\ref{eq18}) the function
$\hat{f}$ depends both on $k\vec{\sigma}$ and on $\vec{\sigma}$ (via
the second and the last terms in (\ref{eq18})) -- a property
postulated in the Landau approach (see Section \ref{IIa}).

We now replace the sum over nodes in (\ref{eq15d}) by a sum over
node degrees
\begin{equation}\label{eq15e}
F(\vec{\sigma},T) = \frac{1}{N}\sum_{i=1}^N \hat{f}(\vec{\sigma},
k_i\vec{\sigma}) = \sum_{k=1}^{k_{max}} P(k) \hat{f}(\vec{\sigma},
k\vec{\sigma}).
\end{equation}
Here $P(k)$ is the density of nodes with degree $k$ (\ref{eq01}) and
$\hat{f}(\vec{\sigma}, k\vec{\sigma})$ represents the contribution
from a single $k$-degree node. Note that $\hat{f}(\vec{\sigma},
k\vec{\sigma})$ actually depends on $\vec{\sigma}$ and $k\sigma$
\begin{equation}
\hat{f}(\vec{\sigma}, k\vec{\sigma}) \equiv \hat{f}(\vec{\sigma},
k\sigma).
\end{equation}
Therefore, we further replace the sum over $k$ in (\ref{eq15e}) by
an integral over $k$ (\ref{eq04}), and introduce $y = k\sigma$ as
the variable of integration.
\begin{equation}\label{eq15f}
F(\vec{\sigma},T) = \sigma^{\lambda-1} \int_{\sigma}^{\infty}P(y)
\hat{f}(\vec{\sigma}, y) {\rm d}y.
\end{equation}
The convergence of the integral (\ref{eq15f}) for large $y$ can be
derived from the asymptotic behavior of the function $I_{\nu}(z)$
\cite{Abramowitz}:
\begin{equation}\label{eq20a}
I_{\nu}(z)\sim \frac{e^z}{\sqrt{2\pi z}}, \hspace{2em}z\to\infty.
\end{equation}
Namely, substituting (\ref{eq20a}) into (\ref{eq18}) one finally
arrives at
\begin{equation}\label{eq20}
\hat{f}(\vec{\sigma}, y) \sim y,\hspace{2em}y\to\infty.
\end{equation}
The last estimate, together with the power law behavior (\ref{eq01})
proves the convergence of the expression for the free energy
(\ref{eq15d}) for $\lambda>2$.

The behavior of $\hat{f}(\vec{\sigma}, y)$ for small $y\to0$ and a
small anisotropy parameter $u/T\ll1$ is characterized by the
smallest term of the Bessel function expansions \cite{Abramowitz} in
\begin{equation}\label{eq20b}
I_{\nu}(z) = \Big(\frac{z}{2}\Big)^{\nu}
\sum_{q=0}^{\infty}\frac{(z^2/4)^q}{k!\, \Gamma(\nu+q+1)},
\end{equation}
where $\Gamma(\rho)$ is Euler gamma function. Now, substituting
(\ref{eq20b}) into (\ref{eq18}) one arrives at:
\begin{eqnarray}\nonumber
\hat{f}(\vec{\sigma}, y) = f_0 + \frac{1}{2}J \sigma y -
\frac{1}{4}\frac{(JL)^2}{T}y^2\\\label{eq19} +
\frac{1}{64}(1-\frac{49}{8}\frac{u
L^4}{T})\frac{(JL)^4}{T^3}y^4\\\nonumber +\frac{1}{384}u
L^4\frac{(JL)^4}{T^4}\frac{\sigma_1^4+\sigma_2^4}{\sigma^4}y^4+\ldots
\end{eqnarray}
with
\begin{equation}\label{eq20c}
f_0 = -T\ln\big(2\pi L\big) + \frac{3}{4}uL^4.
\end{equation}
We will perform the integration in (\ref{eq15f}) using the expansion
of $\hat{f}(\vec{\sigma}, y)$ (\ref{eq19}) and its asymptotics
(\ref{eq20}) at $y\to\infty$.

Those terms of the expansion (\ref{eq19}) that are well behaved with
respect to the integration in (\ref{eq15f}) may be easily
integrated. These are the terms, in which $y^{\mu}$ appears with
$\mu<\lambda-1$. The integration of the remainder of the series
(\ref{eq19}) (let us denote it as $\hat{g}(\vec{\sigma}, y)$) needs
some special care. Using the asymptotic behavior of
$\hat{g}(\vec{\sigma}, y)$ at small and large values of $y$, the
integration is to be performed in the same way as for $g(\vec{x},
y)$ (see Section \ref{IIa}), to obtain the free energy as described
above.

To complete the calculations we now pass from the average spin
$\vec{\sigma}$ per nearest neighbor to the mean spin
$\langle{\vec{s}}\rangle$ of a node. Solving the self-consistency
equation (\ref{eq13b}) for $\langle{\vec{s}}\rangle$ one finds in a
linear approximation in $\vec{\sigma}$ and $u$
\begin{equation}\label{eq20d}
\langle{\vec{s}}\rangle = \frac{J\langle{k}\rangle
L^2}{2T}\vec{\sigma}.
\end{equation}
Substituting (\ref{eq20d}) into (\ref{eq19}) one finally obtains the
free energy density as
\begin{eqnarray}\nonumber
f(\langle\vec{s}\rangle, k\langle\vec{s}\rangle) = f_0 +
\frac{2T^2}{J\langle{k}\rangle^2L^4}k\langle\vec{s}\rangle^2 -
\frac{T}{\langle{k}\rangle^2 L^2}k^2\langle\vec{s}\rangle^2
\\\label{eq20e} + \frac{T}{4\langle{k}\rangle^4L^4}(1-\frac{49}{8}\frac{u
L^4}{T})k^4\langle\vec{s}\rangle^4\\\nonumber +\frac{u k^4}{24
\langle{k}\rangle^4}(\langle{s_1}\rangle^4+\langle{s_2}\rangle^4)+\ldots
\end{eqnarray}
Note, that taking into account higher order corrections in
(\ref{eq20d}) does not change the free energy at critical point.

The expression (\ref{eq20e}) serves as an example for a microscopic
interpretation of the phenomenological Landau free energy
$\Phi(\vec{x},T)$ (\ref{eq04}), (\ref{eq06}). Indeed, the two
component order parameter $\vec{x}$ in (\ref{eq06}) may be
interpreted as the two-component mean spin (magnetization) per site
$\langle\vec{s}\rangle$ in (\ref{eq20e}). The remaining
phenomenological Landau parameters may be found by direct comparison
of the expression (\ref{eq06}) and (\ref{eq20e}). In this way, the
value of $f_0$ in (\ref{eq06}) has a microscopic representation in
terms of (\ref{eq20c}), while the coefficients $a_i$ read:
\begin{equation}\label{eq001}
a_0=0,\hspace{1em}
a_1=\frac{2T^2}{J\langle{k}\rangle^2L^4},\hspace{1em}
a_2=-\frac{T}{\langle{k}\rangle^2L^2}.
\end{equation}

Recall, that in the frames of the Landau approach the assumption
$a_0=0$ leads to the dependence of $T_c$ on
$\langle{k^2}\rangle/\langle{k}\rangle$. Now we find the expression
for $T_c$ in the microscopic model as:
\begin{equation}\label{eq002a}
T_c=\frac{JL^2}{2} \frac{\langle{k^2}\rangle}{\langle{k}\rangle}.
\end{equation}
The values for the other coefficients in the Landau expansion are as
follows
\begin{equation}\label{eqb01}
b_i=c_i=0,\hspace{1em}i = 0\ldots 3,
\end{equation}
\begin{equation}\label{eqb02}
b_4=\frac{T}{4\langle{k}\rangle^4L^4}(1-\frac{49}{8}\frac{u
L^4}{T}),\hspace{1em}c_4=\frac{u}{24\langle{k}\rangle^4}.
\end{equation}

Moreover, the microscopic approach allows us to estimate the
temperature dependence of the proportionality coefficients in the
free energy expressions. The latter is of primary importance for the
case $2<\lambda\leq3$, when the critical temperature diverges. Then
\begin{equation}\label{eq008}
C,C^{'}\sim T^2,\hspace{1em} D,D^{'}\sim T, \hspace{1em} E,E^{'}\sim
T^0.
\end{equation}

In the following, we pass to a more detailed analysis of the Landau
free energy (\ref{eq04}).

\section{Phase diagrams}\label{III}

Having determined the behavior of the free energy (\ref{eq04}), let
us investigate the stable states of the system. The latter may be
found from the minimization of the free energy. The condition of
stationarity requires the first derivatives of the free energy to
vanish
\begin{equation}\label{eq0014}
\frac{\partial \Phi(\vec{x},T)}{\partial x_1}=0, \hspace{2em}
\frac{\partial \Phi(\vec{x},T)}{\partial x_2}=0.
\end{equation}
The stationarity point is a minimum if both eigenvalues of the
matrix of second derivatives
\begin{equation}\label{eq0015}
\omega_{\mu\nu}=\frac{\partial^2 \Phi(\vec{x},T)}{\partial
x_{\mu}\partial x_{\nu}}, \hspace{2em} \mu,\nu=1,2
\end{equation}
are positive. This condition may also be written as
\begin{equation}\label{eq0016}
{\rm Re}(\omega_{\mu\mu})>0,\hspace{1em}
\det(\omega_{\mu\nu})>0,\hspace{1em}\mu,\nu=1,2.
\end{equation}
From a physical point of view, the minimum of the free energy
requires positive isothermal susceptibilities.

In the following we consider the stable states of the system with
coupled order parameters for the relevant ranges of the exponent
$\lambda$, discussed for the generalized Landau free energy.

\subsection{Case $\lambda>5$}

\begin{figure}
\centerline{
\includegraphics[width=60mm]{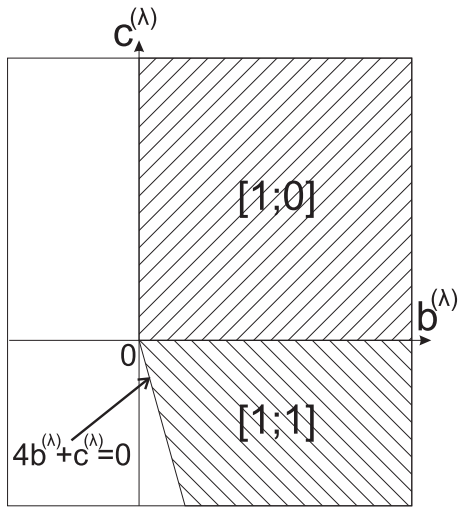}}
\caption{The phase diagram for the coupled two component order
parameter model (\ref{eq02}) on a complex scale-free network. The
picture shows what type of ordering is realized in the different
phases depending on the free energy parameters $b^{(\lambda)}$ and
$c^{(\lambda)}$. The blank part of the phase diagram corresponds to
absence of a stable phase. An ordered phase exists only if
$b^{(\lambda)}>0$. The sign of the coefficient $c^{(\lambda)}$
separates two phases. Namely, positive values of $c^{(\lambda)}>0$
correspond to  phases with only one non-zero order parameter
component ($[1,0]$ or $[0,1]$); negative values $c^{(\lambda)}<0$
that satisfy condition ($4b^{(\lambda)}+c^{(\lambda)}>0$)
corresponds to the ordered phase $[1,1]$, where both order
parameters to have the same non-zero value.}
 \label{fig1}
\end{figure}

For $\lambda>5$ the system is described by the Landau free energy
(\ref{feg5}), whereas the type of the ordering below $T_c$ depends
of the interplay between the fourth-order couplings. If
$c^{(\lambda)}>0$ and $b^{(\lambda)}>0$, the system is characterized
by order parameter components
\begin{equation}\label{eq0018}
x_1 = \sqrt{\frac{a}{b^{(\lambda)}}} (T_c - T)^{\beta}, \hspace{1em}
x_2=0.
\end{equation}
If $c^{(\lambda)}<0$ and $4b^{(\lambda)}+c^{(\lambda)}>0$,  both
order parameters have the same value
\begin{equation}\label{eq0019}
x_1 = x_2 = \sqrt{\frac{2a}{4b^{(\lambda)}+c^{(\lambda)}}} (T_c -
T)^{\beta},
\end{equation}
with $\beta = 1/2$. Here and below we do not write explicitly one
more solution $x_1 = 0, x_2 \neq 0$ which is symmetric to
(\ref{eq0018}) and which is stable under the same conditions as the
solution $x_1 \neq 0, x_2 = 0$. The resulting phase diagram is shown
in Fig. \ref{fig1}.
The blank parts of the phase diagram correspond to cases where no
stable state exists. For these values of $a$, $b^{(\lambda)}$ and
$c^{(\lambda)}$ the condition of stability of the thermodynamic
potential cannot be satisfied (i.e. the asymptotics $\Phi(\vec{x},T)
\to\infty$ for $|\vec{x}|\to\infty$ do not hold), therefore the
system is undefined for this range of parameters.

\subsection{Case $\lambda=5$}

If the exponent $\lambda$ is at its marginal value $\lambda=5$, the
free energy is described by (\ref{fe5}). For temperatures below
$T_c$ stable states exist only if $b^{(\lambda)}>0$. For
$c^{(\lambda)}>0$, the system is described near the critical point
$T\to T_c$ by an ordered phase with
\begin{equation}\label{eq22}
x_1 \approx\sqrt{\frac{2a}{b^{(\lambda)}}}
\frac{(T_c-T)^{\beta}}{[\ln{(T_c-T)^{-1}}]^{1/2}} ,\hspace{1em} x_2
= 0.
\end{equation}
If $-4b^{(\lambda)}<c^{(\lambda)}<0$, the ordered phase at $T\to
T_c$ is characterized by the order parameters
\begin{equation}\label{eq23}
x_1 = x_2 \approx 2\sqrt{\frac{a}{4b^{(\lambda)}+c^{(\lambda)}}}
\frac{(T_c - T)^{\beta}}{[\ln{(T_c - T)^{-1}}]^{1/2}}.
\end{equation}
Note that expressions (\ref{eq22}), (\ref{eq23}) represent
approximate solutions of Eq. (\ref{eq0014}), which is transcendental
when considering the free energy (\ref{fe5}). Both phases are
characterized by the same value of the critical exponent
$\beta=1/2$. For other values of $b^{(\lambda)}$ and $c^{(\lambda)}$
the free energy (\ref{fe5}) does not lead to any equilibrium stable
state, as in the  case  $\lambda>5$. These results are also depicted
in the phase diagram in Fig.\ref{fig1}.

\subsection{Case $3<\lambda<5$}
For degree distributions governed by an exponent in the range
$3<\lambda<5$, the system is described by the free energy
(\ref{fe35}). Below $T_c$ stable states exist only if
$b^{(\lambda)}>0$. Namely, there are two stable phases, one with
\begin{equation}\label{eq25}
 x_1 = \Big(\frac{4a}{(\lambda-1)b^{(\lambda)}}\Big)^{\frac{1}{\lambda-3}} (T_c-T)^{\beta},\hspace{1cm} x_2 = 0,
\end{equation}
and a second one with
\begin{equation}\label{eq26}
x_1 = x_2
=\frac{1}{\sqrt{2}}\Big(\frac{16a}{(\lambda-1)(4b^{(\lambda)}+c^{(\lambda)})}\Big)^{\frac{1}{\lambda-3}}
(T_c - T)^{\beta}
\end{equation}
where $\beta=\frac{1}{\lambda-3}$. The regions where these states
are realized are shown in the phase diagram Fig.\ref{fig1}. If
$c^{(\lambda)}>0$ the system is in the stable state (\ref{eq25}).
Otherwise for negative $c^{(\lambda)}<0$ and
$4b^{(\lambda)}+c^{(\lambda)}>0$ the system is described by the
order parameter components (\ref{eq26}). As observed earlier for
larger values of $\lambda$, the stability conditions (\ref{eq0014}),
(\ref{eq0016}) cannot be satisfied for other values of
$b^{(\lambda)}$ and $c^{(\lambda)}$ parameters in the free energy
(\ref{fe35}).

We conclude that if $\langle{k^4}\rangle$ diverges but
$\langle{k^2}\rangle$ is finite ($3<\lambda\leq5$), the critical
behavior differs from the classical mean-field behavior.
Furthermore, if $\lambda=5$, logarithmic corrections appear, and if
$3<\lambda<5$, the critical exponents are functions of $\lambda$.
Note, that for all values of $\lambda>3$ considered above there
exists a finite critical temperature. This will not be the case for
the values of $\lambda$ considered below.

\subsection{Case $2<\lambda\leq3$}

When the exponent $\lambda$ is in the range $2<\lambda\leq 3$ and
the second moment $\langle{k^2}\rangle$ of the node degree
distribution (\ref{eq01}) diverges, one may infer from (\ref{eq08})
that the order-disorder phase transition does not occur at any
finite temperature. Taking into account that at $T=0$ the system is
ordered, the system keeps order at any finite temperature, as has
been confirmed for the Ising model on the infinite size
Barabasi-Albert scale-free network \cite{Aleksiejuk02,Bianconi02}.

{\em In the case $\lambda = 3$}, the free energy is given by
Eq.(\ref{fe3}). If the parameter $D$ is positive as follows from
Eqs. (\ref{eq01}), (\ref{eq006}), (\ref{eq001}), the system is
always ordered. The type of order found depends on the parameter
$E$. If $E$ is positive only one order parameter has nonzero value
\begin{equation}\label{eq30a}
x_1 = e^{-\frac{2C+D}{2D}}, \hspace{1em} x_2=0.
\end{equation}
For negative values of $E$ both order parameters are nonzero and
have equal value
\begin{equation}\label{eq30b}
x_1 = x_2 = \frac{1}{\sqrt{2}}e^{-\frac{4C+2D+E}{4D}}.
\end{equation}
 The high-temperature
dependence of both order parameters in view of (\ref{eq008}) follows
\begin{equation}
x_1,x_2 \sim e^{-\eta T}
\end{equation}
with some constant $\eta$ determined by the coefficients of the high
temperature behavior of $C$ and $D$ in Eq. (\ref{eq008}).

{\em In the case $2< \lambda < 3$} the system is described by the
free energy (\ref{fe23}). Assuming the anisotropy parameter to be
small ($E^{'}\ll D^{'}$) one finds stable states of the system. If
$C^{'}>0$ (corresponding to (\ref{eq01}), (\ref{eq006}),
(\ref{eq001})) and $D^{'}<0$, the system is always ordered and the
type of order depends on the sign of the anisotropy parameter
$E^{'}$. Namely, if $E^{'}>0$, the ordered phase is characterized by
\begin{equation}
x_1=\Big(\frac{2}{\lambda-1}\Big)^{\frac{1}{\lambda-3}}\Big(-\frac{C^{'}}{D^{'}}\Big)^{\frac{1}{\lambda-3}},\hspace{1em}x_2=0.
\end{equation}
If $E^{'}<0$, both order parameters are nonzero with
\begin{equation}
x_1=x_2=\Big(\frac{2^{\frac{9-\lambda}{2}}}{\lambda-1}\Big)^{\frac{1}{\lambda-3}}
\Big(-\frac{C^{'}}{4D^{'}+E^{'}}\Big)^{\frac{1}{\lambda-3}}.
\end{equation}

Taking into account the high temperature dependence of $C^{'}$ and
$D^{'}$ (\ref{eq008}), the temperature dependencies of the non-zero
order parameters for $T\to\infty$ can be found as:
\begin{equation}\label{eq35}
x_1, x_2 \sim T^{-\frac{1}{3-\lambda}}.
\end{equation}
This corresponds to the obtained for the scalar theory results
\cite{Leone02,Dorogovtsev02b}. As one may expect, for all
$2<\lambda\leq3$ both $x_1$ and $x_2$ vanish only at infinitely
large temperature.

\section{Reaction of the system to an external action}\label{IV}

\subsection{Isothermal susceptibilities}

In the case of two order parameters, the behavior of the system in
an external field is described by two quantities. The longitudinal
susceptibility $\chi_{\parallel}$ describes the reaction of the
system to the external field applied along the order parameter
direction. In turn, $\chi_{\perp}$ describes the reaction to a
transverse external field.

In the disordered state and in absence of an external field the
system is isotropic and therefore there is no difference between
$\chi_{\parallel}$ and $\chi_{\perp}$. In the general case the
susceptibility matrix
 $\chi_{\mu \nu } =(\partial x_\mu /\partial h_\nu )|_T$
(see e.g. \cite{Itzykson89})
\begin{equation}\label{eq36}
\chi_{\mu \nu} = \delta_{\mu \nu}\chi_{\parallel} + (1-\delta_{\mu
\nu})\chi_{\perp}, \hspace{1em} \mu, \nu=1,2
\end{equation}
depends on both $\chi_{\parallel}$ and $\chi_{\perp}$ which may be
found as the inverse eigenvalues of the matrix  of second order
derivatives of the free energy (\ref{eq0015}). Here $\delta_{\mu
\nu}$ is the Kronecker symbol.

Thus, above the critical temperature $T>T_c$ both susceptibilities
have the same dependence, for all values of $\lambda>3$
\begin{equation}\label{eq37}
\chi_{\parallel} = \chi_{\perp} = \frac{1}{a} (T-T_c)^{-\gamma}
\end{equation}
with the standard mean field critical exponent $\gamma=1$. As
mentioned above, there is no disordered state in a scale free
network of infinite size with $2<\lambda\leq3$.

{\em For all $\lambda>3$} and below the critical temperature both
$\chi_{\parallel}$ and $\chi_{\perp}$ follow a power law with the
mean field critical exponent $\gamma = 1$, as also found for the
case of a scalar order parameter. Furthermore, the absolute value of
the longitudinal susceptibility $\chi_{\parallel}$ coincides with
the susceptibility $\chi$ found for the scalar case
\cite{Goltsev03}. Our results are
 \begin{equation}\label{eq37a}
\chi_{\parallel}=\left\{\begin{array}{ll}
 \frac{1}{2a}(T_c-T)^{-\gamma}, & \lambda>5\\
 \frac{1}{(\lambda-3)a}(T_c-T)^{-\gamma}, & 3<\lambda\leq5.
 \end{array}\right.
\end{equation}
The absolute value of the transverse susceptibility $\chi_{\perp}$
depends on both $\lambda$ and the type of order. So, {\em for
$\lambda>5$}
\begin{equation}\label{eq37b}
\chi_{\perp}=\left\{\begin{array}{lll}
 \frac{2b^{(\lambda)}}{ac^{(\lambda)}}(T_c-T)^{-\gamma}, & {\rm for} &\vec{x}=[1,0]\\
 -\frac{4b^{(\lambda)}+c^{(\lambda)}}{2ac^{(\lambda)}}(T_c-T)^{-\gamma}, & {\rm for} &
 \vec{x}=[1,1].
 \end{array}\right.
\end{equation}
As one may see from (\ref{eq37b}), when the coefficient
$c^{(\lambda)}=0$, and thus the system described by the free energy
(\ref{eq02}) becomes isotropic, the transverse susceptibility
diverges $\chi_{\perp}\to\infty$ for any $T<T_c$. This behavior of
$\chi_{\perp}$ is quite physical and is a consequence of the free
energy symmetry: an infinitely small external field applied in a
direction perpendicular to the order parameter, immediately changes
the order parameter orientation. This is the Goldstone phenomenon,
corresponding to the existence of a soft excitation mode in the
ordered phase  \cite{Itzykson89}.

{\em For $3<\lambda\leq5$}, the transverse susceptibility is given
by
\begin{equation}\label{eq37d}
\chi_{\perp}=\left\{\begin{array}{lll}
 \frac{(\lambda-1)b^{(\lambda)}}{2ac^{(\lambda)}}(T_c-T)^{-\gamma}, & {\rm for} & \vec{x}=[1,0]\\
 -\frac{(\lambda-1)(4b^{(\lambda)}+c^{(\lambda)})}{8ac^{(\lambda)}}(T_c-T)^{-\gamma}, & {\rm for} & \vec{x}=[1,1]
 \end{array}\right.
\end{equation}
As discussed above for the case $\lambda>5$, again the transverse
susceptibility diverges for a vanishing parameter $c^{(\lambda)}=0$.

{\em For $\lambda=3$} the longitudinal susceptibility reads
\begin{equation}\label{eq39}
\chi_{\parallel} = \frac{1}{2D}\sim T^{-1}.
\end{equation}
The sign of the transverse susceptibilities depends on the phase:
\begin{equation}\label{ddd1}
\chi_{\perp}=\left\{\begin{array}{ll}
 1/2E, & \vec{x}=[1,0]\\
 -1/2E, & \vec{x}=[1,1].
 \end{array}\right.
\end{equation}

{\em For $2<\lambda<3$} the behavior is similar. The longitudinal
susceptibility follows
\begin{equation}\label{eq39_1}
\chi_{\parallel} = \frac{1}{2(3-\lambda)C^{'}}\sim T^{-2},
\end{equation}
while we have different transverse susceptibility in the two stable
phases
\begin{equation}\label{ddd2}
\chi_{\perp}=\left\{\begin{array}{ll}
 -\frac{\lambda-1}{4}\frac{D^{'}}{C^{'}E^{'}}, & \vec{x}=[1,0]\\
 \frac{\lambda-1}{16}\frac{4D^{'}+E^{'}}{C^{'}E^{'}}, &
 \vec{x}=[1,1].
 \end{array}\right.
\end{equation}

As we learn from the above equations (\ref{eq37}) -- (\ref{eq37d}),
the singularity at the critical point is governed by the mean-field
value of the critical exponent $\gamma=1$. This reproduces the
behavior observed within the Landau theory for systems with a scalar
order parameter on scale-free networks \cite{Goltsev03}. In this
respect passing to a system with a more complicated symmetry does
not appear to modify the universal critical exponents. Note however,
the significant change in other universal quantities, namely the
susceptibility amplitude ratios. Defining the amplitudes for the
susceptibilities by
\begin{equation}\label{eq38}
\chi_i=\left\{\begin{array}{ll}
 \Gamma_{+,i}(T-T_c)^{-\gamma}, & T>T_c\\
 \Gamma_{-,i}(T_c-T)^{-\gamma}, & T<T_c, \hspace{2em} i={\|,\perp}
 \end{array}\right.
\end{equation}
let us compare the amplitude ratios $\Gamma_{+}/\Gamma_{-}$ for
longitudinal and transverse susceptibilities for different phases.
Recall that for a scalar order parameter Landau theory gives
$\Gamma_{+}/\Gamma_{-} = 2$ \cite{Landau}. Correspondingly, for the
free energy (\ref{eq02}) one finds for the longitudinal
susceptibility
\begin{equation}\label{aaa}
(\Gamma_{+}/\Gamma_{-})_\|=2
\end{equation}
while the amplitude ratio for the transverse susceptibility depends
on the type of the ordered phase:
\begin{equation}\label{aab}
(\Gamma_{+}/\Gamma_{-})_{\perp}=\left\{\begin{array}{ll}
 c/2b, & \vec{x}=[1,0]\\
 -2c/(4b+c), & \vec{x}=[1,1],
 \end{array}\right.
\end{equation}
where the notations [1,0] and [1,1] indicate the corresponding
phases. As one can see, the amplitude ratios (\ref{aab}) depend on
the couplings $b$, $c$. For $\lambda>5$ the free energy (\ref{feg5})
is equivalent to that of Eq. (\ref{eq02}) however with coefficients
$b$ and $c$ given by Eq. (\ref{eq003}). Thus the ratio
$\Gamma_{+}/\Gamma_{-}$  attains the same values as for the systems
with free energy (\ref{eq02}). For $\lambda\leq5$ the ratio
$\Gamma_{+}/\Gamma_{-}$ is a function of $\lambda$, similar as it
holds for the order parameter critical exponent $\beta$. So, the
amplitude ratio for the longitudinal susceptibility for all
$3<\lambda\leq5$ reads
\begin{equation}
(\Gamma_{+}/\Gamma_{-})_\| = (\lambda-3).
\end{equation}
For the transverse susceptibilities the ratio depends on the phase,
and respectively, on the values of the coefficients of the free
energy function. Namely, the susceptibility ratios are
\begin{equation}\label{aac}
(\Gamma_{+}/\Gamma_{-})_{\perp}=\left\{\begin{array}{ll}
 \frac{2c^{(\lambda)}}{(\lambda-1)b^{(\lambda)}}, & \vec{x}=[1,0]\\
 -\frac{8c^{(\lambda)}}{(\lambda-1)(4b^{(\lambda)}+c^{(\lambda)})}, &
 \vec{x}=[1,1].
 \end{array}\right.
\end{equation}
The amplitude ratios for the different ranges of $\lambda$ and
phases are summarized in Table \ref{tab01}.

\begin{table*}[ht]
\tabcolsep5.0mm
\begin{center}
\begin{tabular}{cccc}
\hline
$\lambda$       & $(\Gamma_{+}/\Gamma_{-})_{\parallel}$ & $(\Gamma_{+}/\Gamma_{-})_{\perp [1,0]}$& $(\Gamma_{+}/\Gamma_{-})_{\perp [1,1]}$ \\\hline
$\lambda > 5$     &        2        & $c^{(\lambda)}/2b^{(\lambda)}$& $-2c^{(\lambda)}/(4b^{(\lambda)}+c^{(\lambda)})$ \\
$3<\lambda\leq5$  & $\lambda-3$     & $2c^{(\lambda)}/(\lambda-1)b^{(\lambda)}$ & $-8c^{(\lambda)}/(\lambda-1)(4b^{(\lambda)}+c^{(\lambda)})$ \\
\hline
\end{tabular}
\end{center}
\caption{\label{tab01} Amplitude ratios for different ranges of the
$\lambda$ exponent. Second column: amplitude ratio for the
longitudinal susceptibilities; third and fouth columns -- for the
transverse susceptibilities for two different phases, denoted by
$[1,0]$ and $[1,1]$ respectively.}
\end{table*}

Summarizing, we note that for all the range of $\lambda>3$ (where
the critical temperature $T_c$ exists), the behavior of the system
with respect to an external field is governed by a mean-field
critical exponent $\gamma=1$, but the amplitude ratios have
nontrivial forms.

\subsection{Heat capacity}

The heat capacity describes the behavior of the system with respect
to a change in temperature
\begin{equation}\label{eq41}
c_{h}=T\Big(\frac{d S}{d T}\Big)_{h}.
\end{equation}

In the frames of the Landau theory, the coefficient of $|\vec{x}|^2$
in the free energy changes its sign at the critical temperature. The
other coefficients are assumed to be temperature independent. Note,
that for a phase transition on a scale-free network, this assumption
holds also for $\lambda > 3$, whereas for $2<\lambda\leq3$ the
temperature dependencies of the coefficients are described by
(\ref{eq008}). Then one may find the entropy of the system as
\begin{equation}\label{eee}
S=-\big(\frac{\partial \Phi}{\partial T}\big)_{x},
\end{equation}
which for $\lambda>3$ reduces to the simple expression
\begin{equation}\label{eeeg}
S=-\frac{a}{2}|\vec{x}|^2,
\end{equation}
where $|\vec{x}|$ is a function of temperature and external field.
Substituting stable solutions that follow from (\ref{eq0014}) into
(\ref{eee}), one finds the entropy $S$ at fixed external field $h=0$
for each phase. Respectively, the heat capacity may be found by
taking the derivative of the entropy with respect to the temperature
in (\ref{eq41}).

It is known that for a second-order phase transition in simple
magnets the Landau theory predicts a step in the heat capacity at
the critical temperature $T_c$. The behavior of the heat capacity
for a system on a scale-free network is richer. In the standard
mean-field region $\lambda>5$ and below $T_c$ the heat capacity
decreases linearly with the decrease of temperature. At the critical
point the step in the heat capacity is
\begin{equation}\label{eqc1}
\delta c_{h} = \frac{a^2}{2b^{(\lambda)}}T_c.
\end{equation}
Taking into account the microscopic relations (\ref{eq001}),
(\ref{eq002a}), (\ref{eqb02}), one obtains the step in the heat
capacity as follows:
\begin{equation}
\delta c_{h} = 2\frac{\langle{k^2}\rangle^2}{\langle{k^4}\rangle},
\end{equation}
which vanishes for $\lambda\to 5$. For $3<\lambda\leq5$ there is no
step of $c_{h}$ at the critical point. Namely, for $\lambda=5$ we
find the following expressions for the heat capacity at $T<T_c$ in
the phases $[1,0]$ and $[1,1]$, correspondingly:
\begin{widetext}
\begin{equation}\label{eqc01}
c_{h} \approx
\frac{a^2}{b^{(\lambda)}}\frac{T}{\ln(T_c-T)^{-1}},\hspace{1em}{\rm
in\hspace{1ex}phase} \hspace{1em}[1,0],
\end{equation}
\begin{equation}\label{eqc02}
c_{h} \approx
\frac{4a^2}{4b^{(\lambda)}+c^{(\lambda)}}\frac{T}{\ln(T_c-T)^{-1}},\hspace{1em}{\rm
in\hspace{1ex}phase} \hspace{1em}[1,1].
\end{equation}
In the case $3<\lambda<5$ the corresponding formulas read
\begin{equation}\label{eqc03}
c_{h}=\frac{a}{\lambda-3}\Big[\frac{4a}{(\lambda-1)b^{(\lambda)}}\Big]^{2/(\lambda-3)}T(T_c-T)^{(5-\lambda)/(\lambda-3)},
\hspace{1em}{\rm in\hspace{1ex}phase} \hspace{1em}[1,0],
\end{equation}
\begin{equation}\label{eqc04}
c_{h}=\frac{a}{\lambda-3}\Big[\frac{16a}{(\lambda-1)(4b^{(\lambda)}+c^{(\lambda)})}\Big]^{2/(\lambda-3)}T(T_c-T)^{(5-\lambda)/(\lambda-3)},
\hspace{1em}{\rm in\hspace{1ex}phase} \hspace{1em}[1,1].
\end{equation}
\end{widetext}

As one can see from Eqs. (\ref{eqc01})--(\ref{eqc04}), the heat
capacity vanishes as $T\to T_c$ which differs from the case
$\lambda>5$, where the corresponding value at $T_c$ is given by
(\ref{eqc1}). Nevertheless, a maximum of $c_{h}$ is still present
for $3<\lambda\leq5$. Only now, it is shifted from $T_c$ to the
temperature region $T<T_c$. The low temperature behavior of the heat
capacity at $3<\lambda\leq5$ resembles that for $\lambda>5$:
$c_{h}\sim T$. The heat capacity vanishes both at $T=0$ and $T=T_c$
and possesses maximum at an intermediate temperature $0< T_0 < T_c$.
For $\lambda=5$ this temperature coincides with $T_c$ whereas for
lower values of $\lambda$ we find:
\begin{equation}\label{eqc06}
T_0=\frac{\lambda-3}{2}T_c, \hspace{1em}3<\lambda<5.
\end{equation}
Taking into account the explicit calculations of Section \ref{IIa}:
\begin{equation}\label{eqc06a}
T_0=(\lambda-2)\frac{JL^2}{4}, \hspace{1em}3<\lambda<5.
\end{equation}

\begin{figure}
\centerline{
\includegraphics[width=80mm]{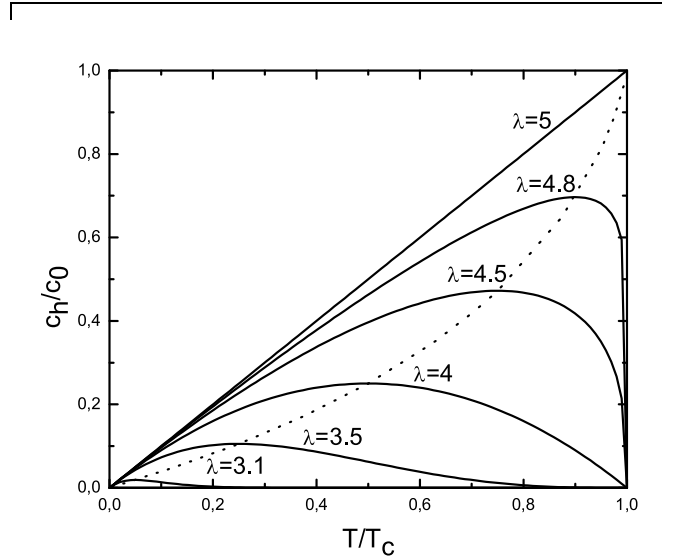}}
\caption{Typical behavior of the heat capacity for different values
of $\lambda$ in the range of $3<\lambda\leq 5$. A dotted curve shows
position of a maximum at temperature $T_0$, see Eq. (\ref{eqc06}).}
 \label{fig2}
\end{figure}

In Fig. \ref{fig2} we show the typical behavior of $c_{h}$ for
different values of $3<\lambda\leq5$. There, we represent Eqs.
(\ref{eqc03}), (\ref{eqc04}) in the form
\begin{equation}\label{eqc06b}
c_{h}= c_0\frac{T}{T_c}\Big(1-\frac{T}{T_c}\Big
)^{(5-\lambda)/(\lambda-3)}
\end{equation}
and plot $c_{h}/c_0$ as a function of a scaled variable $T/T_c$.

As $\lambda$ approaches from above $3$, the critical temperature
increases and becomes infinite for $2<\lambda\leq 3$: the system is
always ordered and the type of ordered phase is governed by signs of
the coefficients $E$, $E^{'}$ in the Landau free energies
(\ref{fe3}), (\ref{fe23}). For both ordered phases we obtain that
the high-temperature behavior of $c_{h}$ is described by
\begin{equation}\label{eq43}
c_{h}\sim\left\{\begin{array}{lc}
 T^2e^{-\zeta T}, & \lambda=3\\
 T^{-\frac{\lambda-1}{3-\lambda}}, & 2<\lambda<3,
 \end{array}\right.
\end{equation}
where $\zeta$ depends on the coefficients defined in Eq.
(\ref{eq008}).

As we have observed for the order parameter and the susceptibility,
the character of the  temperature dependence of the heat capacity
for the system of two coupled order parameters reproduces the one
obtained for a single scalar order parameter
\cite{Dorogovtsev02b,Goltsev03}. Note however the different
amplitudes for this behavior resulting from Eqs. (\ref{eqc03}),
(\ref{eqc04}).

\section{Conclusions}\label{V}

Models that display phase transitions with two coupled order
parameters serve as archetypes to describe the phase behavior in
systems with several possible types of ordering
\cite{Watanabe85,Imry75}. For example, a system may display both
ferromagnetic and antiferromagnetic order with a coupling between
the two. Others may show magnetic and superconducting, ferroelectric
and ferromagnetic, or structural and magnetic order. These models
are known for their rich phase diagrams and non-trivial critical
behavior. Inspired by these observations, this paper investigates
the phase transitions of a corresponding model defined on a
scale-free network. Besides the academic interest, this problem may
have implications for models of opinion formation on social networks
when opinions on different issues may be coupled, as e.g. the
preferences for both a candidate and a political party \cite{note1}.

Our analysis was based on thermodynamic arguments in the spirit of
Landau theory, as suited for the description of phase transitions on
scale free networks \cite{Goltsev03}.  To add a microscopic
background to the phenomenological approach we have also studied a
particular spin Hamiltonian that leads to coupled scalar order
behavior using the mean field approximation. Our results show that
for the scale free networks with a degree distribution governed by
an exponent $\lambda> 2$ the system is characterized by either of
two types of ordering. Either one of the two order parameters is
zero (the $[1,0]$ or the $[0,1]$ phase) or both are non-zero but
have the same value (the $[1,1]$ phase).  Along with the critical
behavior of the scalar order parameter systems on scale-free
networks, the order of the phase transition in the coupled scalar
order parameter system depends on the strength of the node-degree
distribution decay. For rapidly decaying distributions ($\lambda\geq
5$) the second-order phase transition is similar to that described
by usual Landau theory. Nevertheless the new features appear as
$\lambda$ decreases: whereas the magnetic susceptibility (and higher
than the second  derivatives of the free energy with respect to the
magnetic field) remain divergent at $T_c$ for all $3 < \lambda < 5$,
the order of the lowest divergent temperature derivative of the free
energy depends on $\lambda$ \cite{note3}. Namely, as seen from Eqs.
(\ref{eqc03})--(\ref{eqc04}), it is the third order derivative for $
4 < \lambda < 5 $, the fourth order for $3\frac{2}{3} < \lambda <
5$, and so on until it is only the infinite order derivative that
diverges for $\lambda=3$: the order of the phase transition becomes
infinite \cite{Dorogovtsev02b,note2}.

The critical behavior of the model considered gives rise to
non-trivial critical exponents, amplitude ratios and
susceptibilities. While the critical exponents do not differ from
those of a model with a single order parameter on a scale free
network \cite{Goltsev03} there are notable differences for the
amplitude ratios and susceptibilities. Another peculiarity of the
model is that the transverse susceptibility is divergent at all
$T<T_c$,when $O(n)$ symmetry is present. Such behavior  is related
to the appearance of Goldstone modes. It is worth to mention a
peculiarity in the behavior of the specific heat. Whereas for
$\lambda\geq 5$ it has a step at the critical temperature $T_c$,
this step disappears for $\lambda< 5$. The heat capacity vanishes
both at $T=0$ and $T=T_c$ and possesses maximum at an intermediate
temperature $0< T_0 < T_c$.

The phenomena observed serve as  evidence of a rich critical
behavior caused by scale-free properties of the underlying network
structure. An attractive feature for the theoretical analysis of
this behavior is of course that non-trivial effects are found
already in very simple approximations. Natural continuations of our
study will include extensions beyond the mean field approach taking
into account order parameter fluctuations, further, studies of
dynamic processes and in particular the critical dynamics resulting
at or near the critical point. Such studies need to be based on more
detailed information about the structure of the network than the
degree distribution, such as provided by the adjacency matrix or the
network Laplacian (Kirchhoff matrix) e.g. in terms of their
respective eigenvalue spectra \cite{Dorogovtsev03,Jasch04}.

This work was supported by the Fonds zur F\"orderung der
wissenschaftlichen Forschung under Project No. P19583-N20.

\begin{appendix}
\renewcommand{\thesection}{\Alph{section}}%
\renewcommand{\theequation}{\Alph{section}.\arabic{equation}}%
\section{}\label{A}

In order to perform the integration in (\ref{app04}), assume
$4<\lambda<5$. For other values of the exponent $\lambda$ the
calculations can be performed in a similar way. From
Eqs.(\ref{eq11}) and (\ref{eq12a}) we derive the following
asymptotics of $g(\vec{x}, y)$:
\begin{eqnarray}\label{app02}
g(\vec{x},y) &=& \big(b_4 + c_4\frac{x_1^4 +
x_2^4}{|\vec{x}|^4}\big)y^4,\hspace{1em} y\to0\\\label{app03}
g(\vec{x},y) &\sim&y^3,\hspace{1em} y\to\infty \, .
\end{eqnarray}

To analyze Eq. (\ref{app04}) let us define
\begin{eqnarray}\label{app05}
Q_1(\varepsilon, \vec{x}, \lambda) &=&
\int_{\varepsilon}^{\infty}\frac{dy}{y^{\lambda}}g(\vec{x},y), \\
\label{app05a}
 Q_2(\varepsilon, \vec{x}, \lambda) &=&
\int_{\varepsilon}^{|\vec{x}|}\frac{dy}{y^{\lambda}}g(\vec{x},y).
\end{eqnarray}
From the asymptotic behavior (\ref{app03}) one may infer that $Q_1$
is convergent. Assuming that near the critical point the absolute
value of the order parameter $|\vec{x}|\ll1$ is small, we replace
the function $g(\vec{x},y)$ in $Q_2$ by its expansion (\ref{app02})
at small values of $y$ and perform the integration. Then one obtains
\begin{equation}\label{app06}
\int_{|\vec{x}|}^{\infty}\frac{dy}{y^{\lambda}}g(\vec{x},y) =
Q(\vec{x},\lambda) - \big(b_4 + c_4\frac{x_1^4 +
x_2^4}{|\vec{x}|^4}\big)\frac{|\vec{x}|^{5-\lambda}}{5-\lambda}
\end{equation}
where
\begin{equation}\label{app07}
Q(\vec{x},\lambda) = Q_1(\varepsilon, \vec{x}, \lambda) + \big(b_4 +
c_4\frac{x_1^4 +
x_2^4}{|\vec{x}|^4}\big)\frac{\varepsilon^{5-\lambda}}{5-\lambda}.
\end{equation}
Naturally, $Q(\vec{x},\lambda)$ does not depend on $\varepsilon$ (as
$\int_{|\vec{x}|}^{\infty}\frac{dy}{y^{\lambda}}g(\vec{x},y)$ does
not depend on $\varepsilon$), while the dependence of
$Q(\vec{x},\lambda)$ on $\vec{x}$ is reasonably (see the asymptotics
(\ref{app02}) and Eq.(\ref{app07})) to be assumed as follows:
\begin{equation}\label{app08}
Q(\vec{x},\lambda) = v_1 + v_2\frac{x_1^4 + x_2^4}{|\vec{x}|^4}
\end{equation}
where $v_1$ and $v_2$ are some coefficients, in general dependent on
$\lambda$ and the temperature.

Substituting these results into (\ref{app01}), one obtains:
\begin{eqnarray}\nonumber
\int_1^{\infty}dk P(k) g(\vec{x}, k|\vec{x}|) = A
Q(\vec{x},\lambda)|\vec{x}|^{\lambda-1}\\\label{app09} +
\frac{A}{5-\lambda}\big(b_4 + c_4\frac{x_1^4 +
x_2^4}{|\vec{x}|^4}\big)|\vec{x}|^4.
\end{eqnarray}
In the region of $\lambda$ considered ($4<\lambda<5$) near the
critical point the leading term includes a factor
$|\vec{x}|^{\lambda-1}$ and correspondingly $Q(\vec{x},\lambda)$ is
part of the relevant terms of the free energy.

\section{}\label{B}

Here we calculate the partition function (\ref{eq15b}) with
$H_{MF}^i$ described by (\ref{eq15a}). To calculate
\begin{equation}\label{appII_01a}
Z_{MF}=\prod_{i=1}^N Z_{MF}^i= \prod_{i=1}^N{\rm Tr}_i\,
e^{-H_{MF}^i/T}
\end{equation}
we use the following property of the $\delta$-function
\begin{equation}\label{appII_01}
\delta(L-|\vec{s_i}|) = 2L\delta(L^2-|\vec{s_i}|^2)
\end{equation}
and use its Fourier presentation
\begin{equation}\label{appII_02}
\delta(x) = \frac{1}{2\pi {\rm i}}\int_{-{\rm i}\infty}^{{\rm
i}\infty} {\rm d}v_0 e^{v_0x}.
\end{equation}
Then $Z_{MF}^i$ reads
\begin{eqnarray}\label{appII_03}
Z_{MF}^i = \frac{Lz_i}{\pi{\rm i}}\int_{-\infty}^{\infty}{\rm
d}\vec{s_i}\int_{-{\rm i}\infty}^{{\rm i}\infty} {\rm d}v_0
e^{v_0L^2}\\\nonumber \times \prod_{\nu=1}^{2}e^{-u_0s_{\nu,i}^4 -
v_0s_{\nu,i}^2+j_i\sigma_{\nu} s_{\nu,i}}
\end{eqnarray}
with
\begin{equation}\label{appII_04}
z_i = e^{-Jk_i\sigma^2/2T}, \hspace{1em} u_0 = \frac{u}{T},
\hspace{1em} j_i = \frac{J}{T}k_i.
\end{equation}
Now, let us use the representation
\begin{eqnarray}\nonumber
&&\exp\{-u_0s_{\nu,i}^4 - v_0s_{\nu,i}^2+j_i\sigma_{\nu} s_{\nu,i}\}
\\ \label{appII_05}
&&=
\exp\{-\frac{u_0}{j_i^4}\frac{\partial^4}{\partial\sigma_{\nu}^4}\}
\exp\{- v_0s_{\nu,i}^2+j_i\sigma_{\nu} s_{\nu,i}\}
\end{eqnarray}
where
$\exp(-\frac{u_0}{j_i^4}\frac{\partial^4}{\partial\sigma_{\nu}^4})$
is interpreted as
\begin{equation}\label{appII_06}
\exp\{-\frac{u_0}{j_i^4}\frac{\partial^4}{\partial\sigma_{\nu}^4}\}
= 1-\frac{u_0}{j_i^4}\frac{\partial^4}{\partial\sigma_{\nu}^4} +
\frac{1}{2!}\big(\frac{u_0}{j_i^4}\frac{\partial^4}{\partial\sigma_{\nu}^4}\big)^2+\ldots
\end{equation}
Substituting (\ref{appII_05}) into (\ref{appII_03}) one obtains
\begin{eqnarray}\nonumber
&& Z_{MF}^i = \frac{Lz_i}{\pi{\rm i}}
 \Big( \prod_{\nu=1}^2 \exp\{-\frac{u_0}{j_i^4}\frac{\partial^4}{\partial\sigma_{\nu}^4}\} \int_{-\infty}^{\infty}{\rm d}s_{\nu,i}
 \Big)\\\label{appII_07}
&&\times \int_{-{\rm i}\infty}^{{\rm i}\infty} {\rm d}v_0
e^{v_0L^2}\prod_{\nu=1}^{2}\exp\{- v_0s_{\nu,i}^2+j_i\sigma_{\nu}
s_{\nu,i}\}\, .
\end{eqnarray}

To change the order of integration over $S_{\nu,i}$ and $v_0$, we
multiply the integrand with $\exp\{\alpha(L^2 - |\vec{s_i}|^2)\}$,
which is equal to unity due to the constraint. Let us choose
$\alpha$ to be sufficiently large to satisfy $(v_0 +
\alpha)s_{\nu,i}^2 - j_i\sigma_{\nu} s_{\nu,i} > 0$. Then one may
use the Poisson integral
\begin{equation}\label{appII_08}
\int_{-\infty}^{\infty}{\rm d}x e^{-ax^2+bx} =
\sqrt{\frac{\pi}{a}}e^{b^2/4a}
\end{equation}
to obtain
\begin{equation}\label{appII_09}
Z_{MF}^i = \frac{Lz_i}{\pi{\rm i}}
 \int_{\alpha-{\rm i}\infty}^{\alpha+{\rm i}\infty} {\rm d}v
 e^{vL^2} \frac{\pi}{v}
 \prod_{\nu=1}^{2} e^{-\frac{u_0}{j_i^4}\frac{\partial^4}{\partial\sigma_{\nu}^4}}
 e^{j_i^2\sigma_{\nu}^2/4v}.
\end{equation}
Assuming the anisotropy parameter $u$ to be small, and respectively
$u_0\ll1$, we keep only the term linear in $u_0$ in the
(\ref{appII_09}) expansion. Then (\ref{appII_09}) may be written as
\begin{eqnarray}\nonumber
&& Z_{MF}^i = \frac{\pi Lz_i}{\pi{\rm i}}
 \int {\rm d}\omega_i  \exp\{\frac{1}{2}j_iL\sigma(\omega_i+1/\omega_i)\}
 \\ \nonumber && \times \omega_i^{-1} \Big[1
 -u_0L^4\big\{6\frac{\omega_i^{-2}}{(j_iL\sigma)^2} + 6\frac{\omega_i^{-3}}{j_iL\sigma} + \frac{\sigma_1^4 + \sigma_2^4}{\sigma^4}\omega_i^{-4}
 \big\}\Big] \\ \label{appII_10}
\end{eqnarray}
where
\begin{equation}\label{appII_11}
\omega_i = \frac{2Lv}{j_i\sigma}.
\end{equation}
The integration path for the variable $\omega_i$ in the integral in
(\ref{appII_10}) ranges from $2L\alpha/j_i\sigma-{\rm i}\infty$ to
$2L\alpha/j_i\sigma+{\rm i}\infty$.

Using the definition of the modified Bessel function of the first
kind
\begin{equation}\label{appII_12}
I_n(z) = \frac{1}{2\pi {\rm i}}\oint
e^{(z/2)(\omega+1/\omega)}\omega^{-n-1} {\rm d}\omega,
\end{equation}
one may write
\begin{eqnarray}\nonumber
 Z_{MF}^i &=& 2\pi L z_i\Big[I_0(j_iL\sigma) - u_0L^4\big\{
 6\frac{I_2(j_iL\sigma)}{(j_iL\sigma)^2}\\\nonumber
 &+& 6\frac{I_3(j_iL\sigma)}{j_iL\sigma} + \frac{\sigma_1^4 + \sigma_2^4}{\sigma^4}I_4(j_iL\sigma)
 \big\} \Big]. \\ \label{appII_13}
\end{eqnarray}
Substituting (\ref{appII_13}) into (\ref{appII_01a}) one finally
obtains the partition function, and respectively one may find the
free energy per site $F(\vec{\sigma},T) = -T/N\ln Z_{MF}$. Again,
keeping terms linear in $u$, one obtains the free energy
(\ref{eq15d}).
\end{appendix}

\begin{thebibliography}{50}
\bibitem{note1}          Note however that there are different interpretations of the notion
                         of temperature in such formulations, see \cite{soc} .
\bibitem{note3}  Indeed, as one can see from Eqs.
(\ref{eqc03})--(\ref{eqc04}) the heat capacity does not diverge for
$3 < \lambda < 5$.
\end{thebibliography}
\end{document}